\documentclass[onecollarge]{svjour2}       % onecolumn "king-size"
\smartqed  % flush right qed marks, e.g. at end of proof
\usepackage{graphicx}% Include figure files
\usepackage{graphics}
\usepackage{dcolumn}% Align table columns on decimal point
\usepackage{bm}% bold math
\usepackage{float}
\usepackage{caption}
\usepackage{subfigure}
\usepackage{amsmath}
\usepackage{mathrsfs}
\usepackage{amssymb}
\begin{document}

\title{Quantum noise theory for phonon transport through nanostructures }
%\subtitle{many body version of quantum Langevin equation}


\author{Li Wan$^{1*}$   \and Yunmi Huang$^1$ \and Changcheng Huang$^2$%etc.
}


\institute{\at $^1$Department of Physics, Wenzhou University, Wenzhou 325035, P. R. China;\\
$^2$Department of Computer Science, Wenzhou University, Wenzhou 325035, P. R. China;\\
$^*$ \email{lwan@wzu.edu.cn} }

\date{Received: date / Accepted: date}

\maketitle

\begin{abstract}
We have developed a quantum noise approach to study the phonon transport through nanostructures. The nanostructures acting as phonon channels are attached to two phonon reservoirs. And the temperature drop between the two reservoirs drives the phonon transport through the channels. We have derived a quantum Langevin equation(QLE) to describe the phonon transport with the quantum noise originated from the thermal fluctuation of the reservoirs. Within the Markov approximation, the QLE is used to get the thermal conductivity $\kappa$ of the nanostructures and the finite size effect of the $\kappa$ then is studied. In this study, the advantage of the quantum noise approach lays on the fact that no any local temperature needs to be defined for the nanostructures in its non-equilibrium state.   
\end{abstract}

\section{Introduction}
\label{intro}
The development of nano technologies scales electronic devices down to mesoscopic size. Similar to the electrons showing their peculiar behaviors in the nano-structured devices, thermal transport in the structures has been confirmed to have finite size effects by theories and experiments~\cite{siemens,koh,highland,pop,sellan,Dhar,lepri,Wang,gang}. Especially, the thermal conductivity $\kappa$ of the structures decreases with the decreasing of the structure size. Nowadays, study of the finite size effects of the thermal transport has became an important issue and has a significant application in the development of nano technologies~\cite{Dhar,lepri}. For these nanostructures fabricated by dielectric or semiconductor materials, the thermal transport is through lattice interaction rather than by electrons. Various methods have been proposed to study the finite size effect of the thermal transport through the lattice interaction~\cite{Gallavotti,Dhar1,Rieder,Matsuda,lepri1,narayan,lepri2}. Especially, a quantum Langevin equation (QLE) for the lattice vibration in the real space has been obtained, where displacements and momenta of the lattices are used as operators in the equation~\cite{Dhar}. In those real space models, local temperatures have to be defined to describe the energy flow through the lattices. In order to make the temperature definition acceptable in the non-equilibrium systems, the assumption of the coarse-grain-equilibrium(CGE) has to be applied~\cite{Dhar,lepri}. However, for structures with their scales down to a few nanometers, the CGE is expected to be invalid. In order to remove the CGE assumption, we converse the study of the thermal transport in the phonon space, and develop a version of the quantum Langevin equation for the phonon transport by using the quantum noise approach. The creation and annihilation operators of phonons are applied and the quantum noise is originated from the thermal leads. In this paper, we will show the details of the development and then study the finite size effects of the $\kappa$ by using the equation.\\

Generally, the nanostructures studied for the thermal transport are simplified to be an one-dimensional chain with oscillators distributed uniformly in order~\cite{Dhar,lepri}. Two ends of the nanostructures are attached to two thermal leads, which are set at two different temperatures respectively. The drop of temperature between the two leads drives the thermal flow through the chain by the interactions between the oscillators. In order to describe the energy of each oscillator for the thermal flow, local temperatures have to be defined for the oscillators in such non-equilibrium system. It is well known that the temperature actually is a thermodynamical concept for equilibrium systems. Thus, in various models which treat the thermal transport in the real space, the system is assumed to consist of a large number of coarse grains with each grain in its own thermodynamic equilibrium even though the total system is still kept in the non-equilibrium state~\cite{Dhar,lepri}. Such assumption is called as the coarse-grain-equilibrium (CGE), which is invalid for the nanostructures due to the non-equilibrium nature of the coarse grains. However, the CGE assumption can be removed if we convert the study of the thermal transport into the phonon space.\\

Phonons have been well accepted as quasi-particles to describe the lattice vibrations in solids~\cite{ziman}. When the system is in equilibrium state, phonons are stimulated by the temperature and the Bose-Einstein distribution can bridge the phonon density and the temperature. However, in the non-equilibrium cases, it is wrong to define the phonon density at any particular space position due to the fact that phonons are extension lattice waves in the whole system. Therefore, it is meaningless to define the local temperatures in the non-equilibrium system when the phonon space is used for the study of the thermal transport. And then the CGE assumption is removed naturally.\\

In this study, the two thermal leads are in their own equilibrium states and are considered as two phonon reservoirs. The Hamiltonian of the nanostructures can be written in the terms of phonon. The phonon modes of the nanostructures are discrete and determined by the structure size of the system~\cite{wan}. In this way, the finite size effect then is involved by the discrete phonon modes. By coupling phonons of the system and phonons of the reservoirs, the temperature drop between the two reservoirs drives the phonon transport through the nanostructures. The thermal fluctuation of the reservoirs acts as quantum noise, which is used to derive the QLE. The QLE has been developed for electron transport, but it is still lack for the phonon transport~\cite{sun}. In this work, the phonon version of the QLE is obtained.

\section{Quantum Langevin Equation}
\label{sec1}
\subsection{Hamiltonian}
\label{sec2}
The total Hamiltonian $H$ of the full model consists of two parts $H=H_0+H_I$. Here, $H_0$ is the Hamiltonian of the full model without the coupling between the reservoirs and the system. And $H_I$ is the interaction Hamiltonian for the coupling in the model. The $H_0$ and $H_I$ read
\begin{equation}
\label{Ham}
\begin{split}
H_0&=\sum_p \hbar \omega_p^L a^+_p a_p+ \sum_k \hbar \omega_k b^+_k b_k+\sum_q \hbar \omega_q^R c^+_q c_q, \\
H_I&=i\hbar  \sum_{p,k}\xi _{p,k}a^+_pb_k+i\hbar  \sum_{q,k}\eta _{q,k}c_q^+b_k-i\hbar \sum_{p,k}\xi _{p,k}b_k^+a_p-i\hbar \sum_{q,k}\eta _{q,k}b_k^+c_q.
\end{split}
\end{equation}
On the right hand side of the $H_0$, the first and the third terms are the Hamiltonians of the left and right reservoirs respectively, which can be identified by the superscripts $L$ or $R$ of the phonon frequency $\omega$. The second term is the Hamiltonian of the system. The annihilation operators of phonons have been denoted by $a$, $b$ and $c$ for the left reservoir, the system and the right reservoir, respectively. The corresponding create operators are then denoted by  $a^+$, $b^+$ and $c^+$. The subscripts in the Hamiltonians represent the phonon modes. On the right hand side of the $H_I$, $\xi$ is the damping parameter coupling the left reservoir and the system, and $\xi_{p,k}$ is for the coupling between the $p^{th}$ phonon mode and the $k^{th}$ mode. $\eta$ is the coupling between the right reservoir and the system, which has a similar notation of $\xi$.  It should be noted that the phonon modes of the system are selected by the finite size effect and take the values of $\Lambda=\frac{2\pi l}{MD}$ with $M$ the oscillator number and $D$ the lattice parameter of the system~\cite{wan}. The $l$ takes only the integers in the range from $-\frac{M}{2}$ to $\frac{M}{2}$. The finite size effects on the $\kappa$ then can be realized through the number $M$. For the reservoirs, the sizes are regarded to be infinitely long and the temperatures of the reservoirs are not influenced by the system during the phonon transport.

\subsection{Equation of motion}
By using the Heisenberg equation, we get the following equations of motion directly from the Hamiltonians in eq.(\ref{Ham})
\begin{equation}
\label{eqofmot}
\begin{split}
\dot{a}_p&=-i\omega_p^L a_p+\sum_k \xi_{p,k}b_k,\\
\dot{b}_k&=-i\omega_k b_k-\sum_p\xi_{p,k}a_p-\sum_q \eta_{q,k}c_q\\
\dot{c}_q&=-i\omega_q^R c_q+\sum_k \eta_{q,k}b_k.
\end{split}
\end{equation}
The dot on the top of the annihilation operators means the time derivative of the operators. The formal solutions of $a_p$ and $c_q$ read
\begin{equation}
\label{formsolution}
\begin{split}
a_p&=e^{-i\omega^L_p t}a_p(0)+\sum_k \xi_{p,k}\int_0^t dt' e^{-i\omega_p^L(t-t')}b_k(t'),\\
c_q&=e^{-i\omega^R_p t}c_q(0)+\sum_k \eta_{q,k}\int_0^t dt' e^{-i\omega_q^R(t-t')}b_k(t').
\end{split}
\end{equation}
Here, $a_p(0)$ and $c_q(0)$ are referred to as the operators at the instant when the coupling between the system and the reservoirs was just switched on. Multiplying both sides of the equations in eq.(\ref{formsolution}) by the damping parameters, we obtain 
\begin{equation}
\label{noise1}
\begin{split}
\sum_p \xi _{p,k}a_p&=\sum_p \xi_{p,k}e^{-i\omega^L_p t}a_p(0)+\sum_{p,k'}\xi_{p.k}\xi_{p,k'}\int_0^tdt' e^{-i\omega_p^L(t-t')}b_{k'}(t'),\\
\sum_q \eta _{q,k}c_q&=\sum_q \eta_{q,k}e^{-i\omega^R_q t}c_q(0)+\sum_{q,k'}\eta_{q.k}\eta_{q,k'}\int_0^tdt' e^{-i\omega_q^R(t-t')}b_{k'}(t').
\end{split}
\end{equation}
Substitute eq.(\ref{noise1}) into the equation of $\dot{b}_k$ in the eq(\ref{eqofmot}) to get the equation of motion
\begin{equation}
\label{eqofmotb}
\begin{split}
\dot{b}_k&=-i\omega_k b_k\\
&-\sum_p \xi_{p,k}e^{-i\omega^L_p t}a_p(0)-\sum_{p,k'}\xi_{p.k}\xi_{p,k'}\int_0^tdt' e^{-i\omega_p^L(t-t')}b_{k'}(t')\\
&-\sum_q \eta_{q,k}e^{-i\omega^R_q t}c_q(0)-\sum_{q,k'}\eta_{q.k}\eta_{q,k'}\int_0^tdt'e^{-i\omega_q^R(t-t')}b_{k'}(t').
\end{split}
\end{equation}

\subsection{Markov approximation}
For simplicity, we note 
\begin{equation}
\mathcal{L}_k=-\sum_p \xi_{p,k}e^{-i\omega^L_p t}a_p(0);~~~~\mathcal{R}_k=-\sum_q \eta_{q,k}e^{-i\omega^R_q t}c_q(0).
\end{equation}
The $\mathcal{L}_k$ and $\mathcal{R}_k$ are the quantum noise induced by the left and right reservoirs respectively, and act on the $k^{th}$ phonon mode of the system. The temperatures of the left and the right reservoirs are denoted by $T^L$ and $T^R$ respectively. The Bose-Einstein distribution bridges the phonon densities and the temperatures of the reservoirs by
\begin{equation}
\label{BEavg}
\begin{split}
&<a^+_k(0)a_{k'}(0)>=\frac{\delta_{k,k'}}{e^{\hbar \omega_k^L/k_BT^L}-1},~~~~ <a_k(0)a_{k'}^+(0)>=\frac{\delta_{k,k'}}{1-e^{-\hbar \omega_k^L/k_BT^L}},\\
&<c^+_k(0)c_{k'}(0)>=\frac{\delta_{k,k'}}{e^{\hbar \omega_k^R/k_BT^R}-1},~~~~ <c_k(0)c_{k'}^+(0)>=\frac{\delta_{k,k'}}{1-e^{-\hbar \omega_k^R/k_BT^R}}.
\end{split}
\end{equation}
For convenience, we use $\bar{N}(\omega_k, T)$ to denote the distribution $\bar{N}(\omega_k, T)=\frac{1}{e^{\hbar \omega_k/k_BT}-1}$ in the following derivation. The correlation of the $\mathcal{L}_k$ then reads
\begin{equation}
\label{corL1}
<\mathcal{L}_k^+(t)\mathcal{L}_{k'}(t')>=\sum_{p}\xi_{p,k}\xi_{p,k'}e^{-i(\omega^L_{p}t'-\omega^L_pt)}\bar{N}(\omega_p^L, T^L).
\end{equation}
Since the phonon frequencies of the reservoirs are continuous, the sum of the frequencies in eq.(\ref{corL1}) can be transformed in integral, reading
\begin{equation}
\label{corL2}
<\mathcal{L}_k^+(t)\mathcal{L}_{k'}(t')>=\int d \omega^L_p \frac{dD(\omega_p^L)}{d \omega_p^L} \xi_{p,k}\xi_{p,k'}e^{-i(\omega^L_{p}t'-\omega^L_pt)}\bar{N}(\omega_p^L, T^L).
\end{equation}
The factor of $\frac{dD(\omega_p^L)}{d \omega_p^L}$ in the above equation is the density of state of phonons. Now we suggest the Markov Approximation by
\begin{equation}
\label{MarkovapproxL}
\frac{dD(\omega_p^L)}{d \omega_p^L} \xi_{p,k}\xi_{p,k'}=\frac{ \omega_p^L \gamma^L_{k,k'} }{2\pi}.
\end{equation} 
Here, $\gamma^L_{k,k'}$ is the coupling strength between the $k^{th}$ and the $k'^{th}$ phonon modes of  the nanostructures. Physically, the $\gamma^L_{k,k'}$ is realized through the coupling of the system and the left reservoir by $\xi_{p,k}$and $\xi_{p,k'}$. Substituting the eq.(\ref{MarkovapproxL}) into the eq.(\ref{corL2}), we get the correlation of $\mathcal{L}$ as
\begin{equation}
\label{correL}
<\mathcal{L}_k^+(t)\mathcal{L}_{k'}(t')>=\int d \omega^L_p \frac{ \omega_p^L \gamma^L_{k,k'} }{2\pi}e^{-i(\omega^L_{p}t'-\omega^L_pt)}\bar{N}(\omega_p^L, T^L)= \gamma^L_{k,k'} f_L(t-t'),
\end{equation}
with
\begin{equation}
\label{MK1}
 f_L(t-t')=\int d \omega^L_p \frac{ \omega_p^L}{2\pi}e^{-i(\omega^L_{p}t'-\omega^L_pt)}\bar{N}(\omega_p^L, T^L).
\end{equation}
In the above treatment, the Gardiner's consideration must be borrowed that the $\gamma^L$ should drop off at high frequencies, even though it is treated as a constant in the following calculation~\cite{Gardiner}. For simplicity, a cut-off frequency $\omega_c$ is introduced to show the drop-off behavior. Then $\bar{N}(\omega_p^L, T^L)$ can be expanded as $\bar{N}(\omega_p^L, T^L)\approx\frac{k_B T^L}{\hbar \omega_p^L}$ under the condition that the temperature $T^L$ is high enough with $\hbar \omega_c<<k_B T^L$ satisfied. And then the $f_L(t-t')$ recovers $f(t-t')=(k_B T^L/\hbar) \delta(t-t')$ to give a well behaved correlation function of $<\mathcal{L}_k^+(t)\mathcal{L}_{k'}(t')>$. Similarly, we treat the right reservoir like what we have done to the left reservoir, obtaining the correlation
\begin{equation}
\label{correR}
<\mathcal{R}_k^+(t)\mathcal{R}_{k'}(t')>=\int d \omega^R_q \frac{\omega_q^R \gamma^R_{k,k'} }{2\pi}e^{-i(\omega^R_{q}t'-\omega^R_qt)}\bar{N}(\omega_q^R, T^R)= \gamma^R_{k,k'} f_R(t-t').
\end{equation}
with the Markov approximation of
\begin{equation}
\label{MarkovapproxR}
\frac{dD(\omega_q^R)}{d \omega_q^R} \eta_{q,k}\eta_{q,k'}=\frac{ \omega_q^R \gamma^R_{k,k'} }{2\pi},
\end{equation}
and 
\begin{equation}
 f_R(t-t')=\int d \omega^R_q \frac{ \omega_q^R}{2\pi}e^{-i(\omega^R_{q}t'-\omega^R_qt)}\bar{N}(\omega_q^R, T^R).
\end{equation}

\subsection{Damping terms}
With the Markov approximation suggested, we simplify the damping terms in the equation of motion of eq.(\ref{eqofmotb}). The simplification is performed term by term, shown as follows
\begin{equation}
\begin{split}
T1&=\sum_{p,k'}\xi_{p,k}\xi_{p,k'}\int_0^tdt' e^{-i\omega_p^L(t-t')}b_{k'}(t')\\
&= \sum_{k'}\frac{ \gamma^L_{k,k'}}{i}\int_0^tdt' b_{k'}(t')\frac{d}{d t'}\delta(t-t')= \sum_{k'}i \gamma^L_{k,k'} \dot{b}_{k'}(t),
\end{split}
\end{equation}
and
\begin{equation}
\begin{split}
T2&=\sum_{q,k'}\eta_{q.k}\eta_{q,k'}\int_0^tdt' e^{-i\omega_q^R(t-t')}b_{k'}(t')=\sum_{k'}i \gamma^R_{k,k'} \dot{b}_{k'}(t).
\end{split}
\end{equation}
Substituting the damping terms $T_1$ and $T_2$ into the eq.(\ref{eqofmotb}) results in the following simplified equation of
\begin{equation}
\label{simplifiedb}
\dot{b}_k=-i\omega_k b_k+\mathcal{L}_k+\mathcal{R}_k+i \sum_{k'} \Gamma_{k,k'}\dot{b}_{k'},
\end{equation}
with $\Gamma_{k,k'}=\gamma_{k,k'}^L+\gamma^R_{k,k'}$.

\subsection{Quantum Langevin equation}
Define the vectors $b=[\dots b_k \dots]^{Tran}$, $\mathcal{L}=[\dots \mathcal{L}_k \dots]^{Tran}$, and $\mathcal{R}=[\dots \mathcal{R}_k \dots]^{Tran}$ with $Tran$ meaning the transpose of the vectors, and define the following matrices
\begin{equation}
A_{k,k'}=\delta_{k,k'}-i \Gamma_{k,k'}, ~~~~~~\Omega_{k,k'}=\omega_k \delta_{k,k'}.
\end{equation}
Then the set of equations of eq(\ref{simplifiedb}) with all the wave-vectors $k$ can be casted in matrix form as
\begin{equation}
\label{Matrixb1}
A \dot{b}=-i \Omega b+\mathcal{L}+ \mathcal{R}.
\end{equation}
Multiplying the inverse of the matrix $A$ on both sides of the eq.(\ref{Matrixb1}), we get
\begin{equation}
\label{Matrixb2}
 \dot{b}=-i A^{-1}\Omega b+A^{-1}\mathcal{L}+ A^{-1}\mathcal{R}.
\end{equation}
It is easy to find a matrix $U$ to diagonalize the coefficient matrix of $A^{-1}\Omega$, leading to a diagonalized matrix $\Xi=U^{-1}A^{-1}\Omega U$. For further derivation, we define the following matrices
\begin{equation}
B=U^{-1}b,~~\mathbb{L}=U^{-1}\mathcal{L}, ~~\mathbb{R}=U^{-1}\mathcal{R},~~\Lambda=U^{-1}A^{-1}U.
\end{equation}
Note that $U$ for the diagonal is not required to be unitary since the coefficient matrix $A^{-1}\Omega$ is a complex Matrix. Then we get the QLE from the eq.(\ref{Matrixb2}) in the matrix form as
\begin{equation}
\label{PVQLE}
\dot{B}=-i \Xi B+\Lambda \mathbb{L}+\Lambda \mathbb{R}.
\end{equation}
In element, it is
\begin{equation}
\dot{B}_k=-i \Xi_k B_k+\sum_{k'}\Lambda_{k,k'} \mathbb{L}_{k'}+\sum_{k'}\Lambda_{k,k'} \mathbb{R}_{k'}.
\end{equation}
The formal solution of the $B_k$ reads
\begin{equation}
\label{formsoluB}
B_k=e^{-i \Xi_k t} B_k(0)+\sum_{k'} \Lambda _{k,k'}\int_0^t d t' e^{-i \Xi_k(t-t')}(\mathbb{L}_{k'}(t')+\mathbb{R}_{k'}(t')).
\end{equation}
It is very interesting to find that the thermal transport of nanostructures is not realized by every independent mode of phonons, but by their collective modes. The collective modes of the phonons are originated from the coupling between the system and the reservoirs, which now are represented by the $B_k$ instead of $b_k$. The last two terms on the right hand side of the eq.(\ref{formsoluB}) act as the quantum noise applied on the quasi-particles $B_k$. In the following section, we will use the phonon version of QLE eq.(\ref{PVQLE}) to study the thermal transport of the nanostructures, and the thermal conductivity $\kappa$ will be obtained.

\section{Thermal transport}
\subsection{Thermal current}
We set $T^R>T^L$ to drive the thermal current from right to left. The total thermal current can be defined as the rate of energy of the right reservoir by
\begin{equation}
\label{current}
I=\sum_q\frac{d(\hbar \omega^R_qc^+_qc_q)}{dt}=\sum_q \hbar \omega^R_q(\dot{c}^+_qc_q+c^+_q\dot{c}_q).
\end{equation}
By using the equation of motion of $c_q$ in eq.(\ref{eqofmot}) and the Markov approximations of eq.(\ref{MarkovapproxL}) and  eq.(\ref{MarkovapproxR}), we simplify the thermal current eq.(\ref{current}) to be 
\begin{equation}
I=\sum_{q,k} \hbar \omega^R_q \eta_{q,k}(e^{-i\omega^R_p t}b_k^+c_q(0)+e^{i\omega^R_p t}c^+_q(0) b_k)-\sum_{k,k'}\hbar \gamma_{k,k'}^R[b_k^+(t)\ddot{b}_{k'}(t)+\ddot{b}_{k'}^+(t)b_{k}(t)],
\end{equation}
with $\ddot{b}$ representing the second time derivative of $b$. The $\ddot{b}$ is originated from the Fourier transformation of an integrand having a factor of $\omega ^2$. We can transform the thermal current in terms of $B_k$, showing
\begin{equation}
\label{thermcur}
I=-i\hbar B^+ F_0\dot{\mathbb{R}}-\hbar B^+ \Upsilon^R \ddot{B}+H.C.
\end{equation}
with the $H.C.$ meaning the Hermitian Conjugate of the terms. Here we have defined the matrices of $F_0=U^+U,~~\Upsilon^R=U^+\gamma^R U$ for convenience. Since the $U$ is not necessarily unitary, the $F_0$ could be non-unit matrix. According to the QLE of eq.(\ref{PVQLE}), taking the second time derivative of $B$ leads to 
\begin{equation}
\ddot{B}=-\Xi \Xi B-i \Xi \Lambda \mathbb{L}-i \Xi \Lambda \mathbb{R}+\Lambda \dot{\mathbb{L}}+\Lambda \dot{\mathbb{R}}.
\end{equation}
By substituting the expression of $\ddot{B}$ into the equation of eq.(\ref{thermcur}), we get the explicit expression of the total thermal current as the following
\begin{equation}
\label{totalthercur}
I= \hbar B^+ F_1 B+i\hbar B^+ F_2 \mathbb{L}+i\hbar B^+ F_2 \mathbb{R}-\hbar B^+ F_3 \dot{\mathbb{L}}-\hbar B^+ (F_3+iF_0) \dot{\mathbb{R}}+H.C.
\end{equation}
with the notations of
\begin{equation}
F_1=\Upsilon^R \Xi \Xi,~~~~F_2=\Upsilon^R \Xi \Lambda,~~~~F_3=\Upsilon^R\Lambda.
\end{equation}
By substituting the eq.(\ref{formsoluB}) the formal solution of $B$  into the eq.(\ref{totalthercur}), and then making the statistical average, we know that the total thermal current actually is determined by the statistical correlations of $<\mathbb{L}^+\mathbb{L}>$, and $<\mathbb{R}^+\mathbb{R}>$. The two reservoirs are regarded to be independent to each other with zero correlations of $<\mathbb{L}^+\mathbb{R}>=0$. In the total current stimulated by the right reservoir, only the terms of $<\mathbb{L}^+\mathbb{L}>$ determine the exact thermal current $J$ flowing from the right reservoir to the left one~\cite{Dhar1}. Thus, we can separate the $J$ from the total thermal current $I$ by $J=J_1+J_2+J_3+H.C.$, with $J_1= \hbar B^+ F_1 B$, $J_2=i\hbar B^+ F_2 \mathbb{L}$, and $J_3=-\hbar B^+ F_3 \dot{\mathbb{L}}$. By using the formal solution eq.(\ref{formsoluB}), the Markov approximations of eq.(\ref{MarkovapproxL}) and eq.(\ref{MarkovapproxR}), and the basic statistical correlations of eq.(\ref{correL}) and eq.(\ref{correR}), we can reach the expression of $J$. In order to show the expression of $J$ much clearly,  we define the following matrices $G_1^L$, $G_2^L$ and $G_3^L$ for the left reservoir with the elements of the matrices as
\begin{equation}
\label{Gmatrices}
\begin{split}
&(G_1^L)_{k,k_1}=(F_1)_{k,k_1}\int d \omega \omega \cdot \frac{1}{ \omega- \Xi_k^*} \cdot \frac{1}{ \omega-\Xi_{k_1}}\bar{N}(\omega, T^L),\\
&(G_2^L)_{k,k_1}=(F_2)_{k,k_1}\int d \omega \omega \frac{1}{ \omega- \Xi_k^*}\bar{N}(\omega,T^L),\\
&(G_3^L)_{k,k_1}=(F_3)_{k,k_1}\int d \omega  \omega \omega\frac{1}{ \omega- \Xi_k^*}\bar{N}(\omega,T^L).
\end{split}
\end{equation}
Here, $\Xi_k^*$ is the conjugate of the $\Xi_k$. Similarly, we define the matrices of $G_1^R$, $G_2^R$ and $G_3^R$ for the right reservoir by replacing the $\bar{N}(\omega,T^L)$ with $\bar{N}(\omega,T^R)$ in the integrals. Then the $J$ of the system in steady state is obtained as
\begin{equation}
\label{thermalcurrnon}
J_s=\frac{\hbar}{2\pi } tr[W U^{-1} \gamma (U^{-1})^+ \Lambda^+]+H.C.
\end{equation}
with $W=(G_1^R-G_1^L) \Lambda+(G_2^R-G_2^L)+(G_3^R-G_3^L)$. And $tr$ means the trace of the matrix. The damping parameters $\gamma^L$ and $\gamma^R$ are assumed to be equal to each other and both have been written as $\gamma$. It can be found that the $J_s$ equals zero when $T^R=T^L$.

\subsection{Small temperature drop}
We denote the temperature drop between the two reservoirs by  $T^R-T^L=\Delta T$. For the case with small temperature drop of $\Delta T \rightarrow 0$, we can approximate the term $\bar{N}(\omega, T^R)-\bar{N}(\omega, T^L)$ by 
\begin{equation}
\label{linearcase}
\bar{N}(\omega, T^R)-\bar{N}(\omega, T^L)=\dfrac{\hbar \omega e^{\hbar \omega/k_BT} \Delta T}{k_BT^2 (e^{\hbar \omega/k_B T}-1)^2}
\end{equation}
with $T=(T^L+T^R)/2$. Then the $J_s$ can be expanded to be linear to $\Delta T$, which reads  
\begin{equation}
J_s=\frac{\hbar ^2 \Delta T}{2\pi k_B T^2} tr[Q U^{-1} \gamma (U^{-1})^+ \Lambda^+]+H.C.
\end{equation}
with $Q=G_1 \Lambda+G_2+G_3$. The matrix $Q$ actually is reduced from the matrix $W$ under the condition of $\Delta T \rightarrow 0$, with the following definitions of the matrices 
\begin{equation}
\begin{split}
&(G_1)_{k,k_1}=(F_1)_{k,k_1}\int d \omega \omega \cdot \frac{1}{ \omega- \Xi_k^*} \cdot \frac{1}{ \omega-\Xi_{k_1}}\dfrac{ \omega e^{\hbar \omega/k_BT} }{ (e^{\hbar \omega/k_B T}-1)^2},\\
&(G_2)_{k,k_1}=(F_2)_{k,k_1}\int d \omega \omega \frac{1}{ \omega- \Xi_k^*}\dfrac{ \omega e^{\hbar \omega/k_BT} }{ (e^{\hbar \omega/k_B T}-1)^2},\\
&(G_3)_{k,k_1}=(F_3)_{k,k_1}\int d \omega  \omega \omega\frac{1}{ \omega- \Xi_k^*}\dfrac{ \omega e^{\hbar \omega/k_BT} }{ (e^{\hbar \omega/k_B T}-1)^2}.
\end{split}
\end{equation}

\subsection{Thermal conductivity}
The thermal conductivity $\kappa$ can be defined by the classical Fourier law
\begin{equation}
J_s=-\kappa \frac{\Delta T}{MD}.
\end{equation}
The minus on the right hand side means the $J_s$ has an opposite direction of the temperature drop $\Delta T$. In the case of $\Delta T\rightarrow 0$, $\kappa$ has a simple expression
\begin{equation}
\label{linearkappa}
\kappa=\frac{\hbar ^2 MD}{2\pi k_B T^2} tr[Q U^{-1} \gamma (U^{-1})^+ \Lambda^+]+H.C..
\end{equation} 
We call this case as the linear case in the following calculation. And for the case with large $\Delta T$ in which the eq.(\ref{linearcase}) is not satisfied, the $\kappa$ has no analytical solution like that in eq.(\ref{linearkappa}). And then the $\kappa$ is functional of both of $T^R$ and $T^L$ instead of $\Delta T$ only. Such case will be claimed as nonlinear case in the following calculation.

\section{Results and discussions}
In this study, the dispersion relation of the system takes the form of $\omega=\omega_0 |\sin (k D/2)|$ with $\omega_0=4\times 10^{12}Hz$ and $D=1nm$ for simplicity. As mentioned, the wave vectors take the values of $\Lambda=\frac{2\pi l}{MD}$ with $l$ the integers in the range from  $-M/2$ to $ M/2$. We set a reference temperature $T_0=\frac{\hbar \omega_0}{k_B}=30.553K$ and $\kappa_0=\frac{\hbar \omega_0^2D}{2\pi T_0}=8.7\times 10^{-21}\frac{W\cdot m}{K}$ for dimensionless normalization. Note that the unit of $\kappa_0$ in our model is for the one dimensional system, not for the bulk case. The coupling parameter $\gamma$ between the system and the reservoirs takes a constant value of $C$. To understand the results obtained in the following calculation, one has to focus on the influence of four parameters on the $\kappa$. The four parameters are the oscillator number $M$ in the chain, the temperature drop $\Delta T$, the temperature of either reservoir $T^L$ or $T^R$, finally the coupling parameter $C$.   

\subsection{linear case}
In this case, the temperature drop is small and the eq.(\ref{linearcase}) is valid with $\Delta T \rightarrow 0$. The $\kappa$ then can be calculated from eq.(\ref{linearkappa}), and shows the finite size effect of the $\kappa$ in fig.1a. 
\begin{figure}[!htb]
\includegraphics[width=3in]{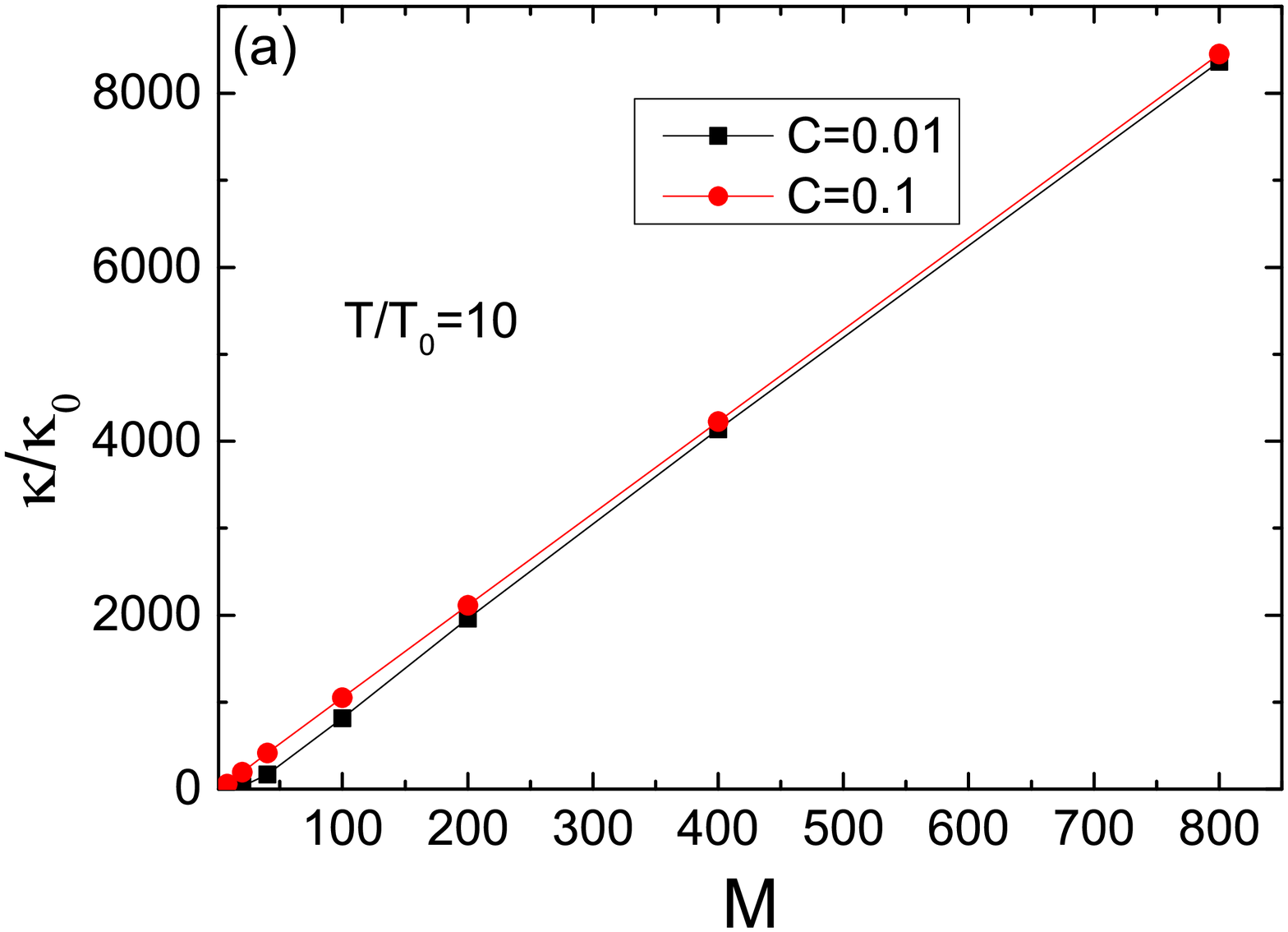}
\includegraphics[width=3in]{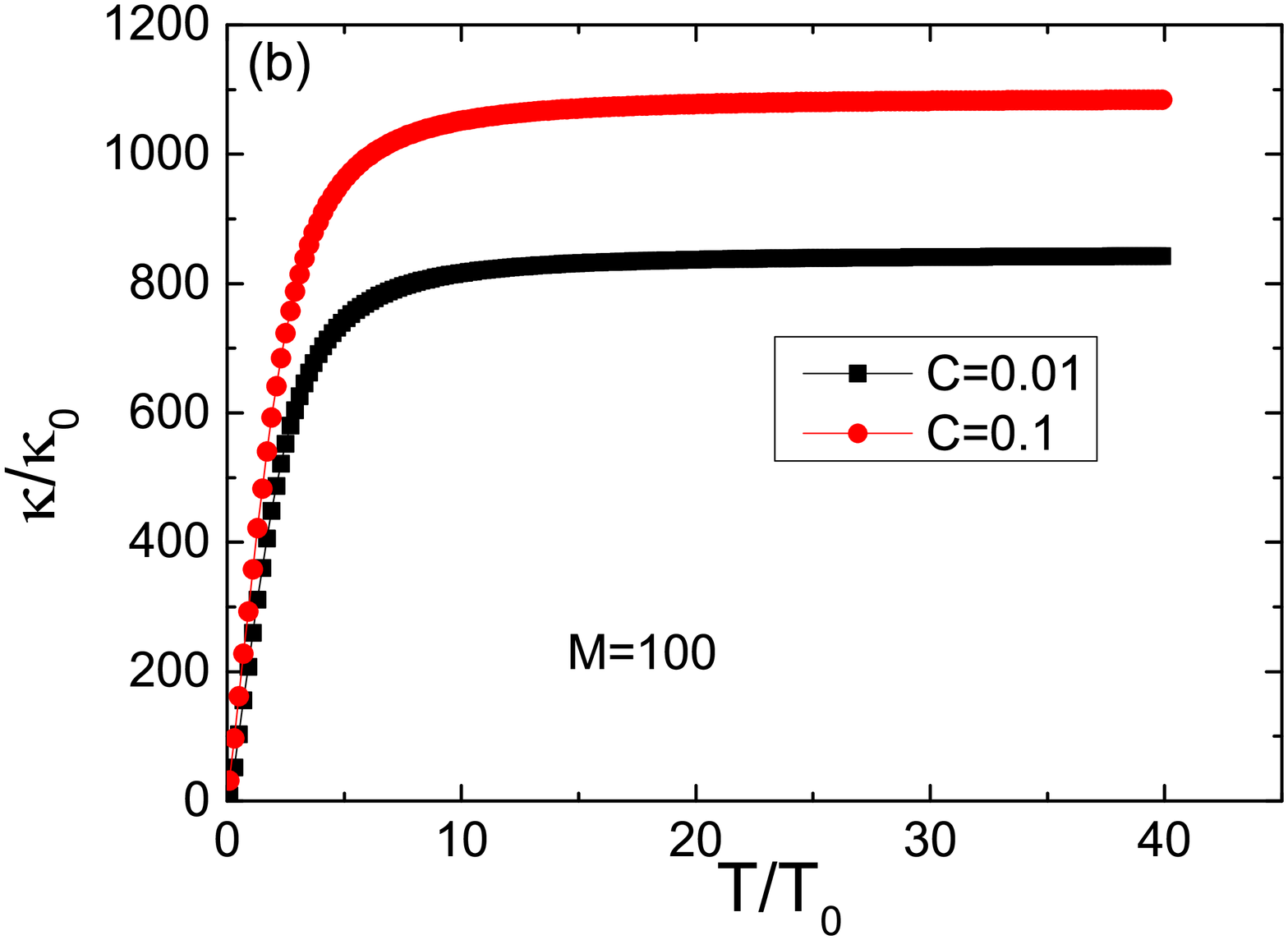}
\caption{linear case for $\kappa$ with a small temperature drop $\Delta T$ between the two reservoirs. The $\kappa_0$ and $T_0$ have been defined for dimensionless normalization. (a) the $\kappa$ is linearly dependent on the oscillator number $M$ of the chain and shows the finite size effect.(b) The $\kappa$ increases with the increasing of $T$ and reaches a constant value for high $T$.}
\end{figure}
In the figure, $\kappa$ decreases with the decreasing of $M$, which is due to fewer phonon channels provided by smaller size of the system for the thermal transport. Such finite size effect of $\kappa$ has been confirmed by the experiments~\cite{Wang,gang}. However, it is not expected that the $\kappa$ will be infinite for an infinitely-increased chain. This is because the disorder of the oscillator chain, as well as the nonlinear interaction between the oscillators, are not considered in our model for simplicity, since we only focus on the thermal transport in nanostructures instead of bulk materials. For the bulk case, the disorder and the nonlinearity will enhance the phonon scatting and lead to a finite and saturate $\kappa$~\cite{wan}. The relation between the $\kappa$ and $M$ can be fitted by $\kappa \sim M^{\alpha}$ to get the exponent $\alpha=1$, which is coincident to the reported result for the harmonic oscillator chain~\cite{Rieder}. The influence of the coupling parameter $C$ on the $\kappa$ has also been studied, and shown in the figure that $\kappa$ decreases with the $C$ decreasing. For the limit case of $C=0$, the thermal transport is switched off. \\

The influence of the $C$ on the $\kappa$ is much more clear in the fig.1b. Larger $C$ means stronger coupling between the system and the reservoirs, contributing more phonons to the thermal transport and a larger $\kappa$. The $T$ dependent behaviors of $\kappa$ have also been revealed by the fig.1b. Lower $T$ stimulates fewer phonon channels for the thermal transport with a smaller $\kappa$. With the $T$ increasing, the $\kappa$ increases due to more and more phonons taking part in the thermal transport. It is known that the function $e^{\hbar \omega/k_B T}$ can be expanded as $1+\hbar \omega/k_B T$ if $\hbar \omega/k_B T<<1$ is satisfied for large $T$. Under such condition, the right hand side of the eq.(\ref{linearcase}) is reduced to be $T$ independent. That means the $\kappa$ reaches a constant with $T$ increasing, which has been clearly shown in the fig.1b. It should be noted here that for too high temperature $T$, the chain may be melted and the $\kappa$ may decrease after a peak. However, such problem is not in the scope of this study. 

\subsection{nonlinear case}
When the eq.(\ref{linearcase}) is invalid for large $\Delta T$, the $\kappa$ has to be calculated numerically by using the eq.(\ref{thermalcurrnon}). Results are shown in fig.2. As revealed in fig.2a, the $\kappa$ increases with the increasing of $\Delta T$. For a larger $\Delta T$, the averaged temperature $T$ of the system is larger when the $T^L$ is fixed. The larger $T$ then produces more phonon channels for the transport and get a larger $\kappa$. For the system with $T^L$ fixed, if the $\Delta T$ is large enough that the $\bar{N}(\omega, T^R)\approx \frac{k_B T^R}{\hbar \omega}$ is satisfied, the $J_s$ linearly dependents on the $\Delta T$ and $\kappa$ then goes to a constant value and is less dependent on $\Delta T$, which has been shown in the fig.2a and fig.2b. 
\begin{figure}[!htb]
\includegraphics[width=3in]{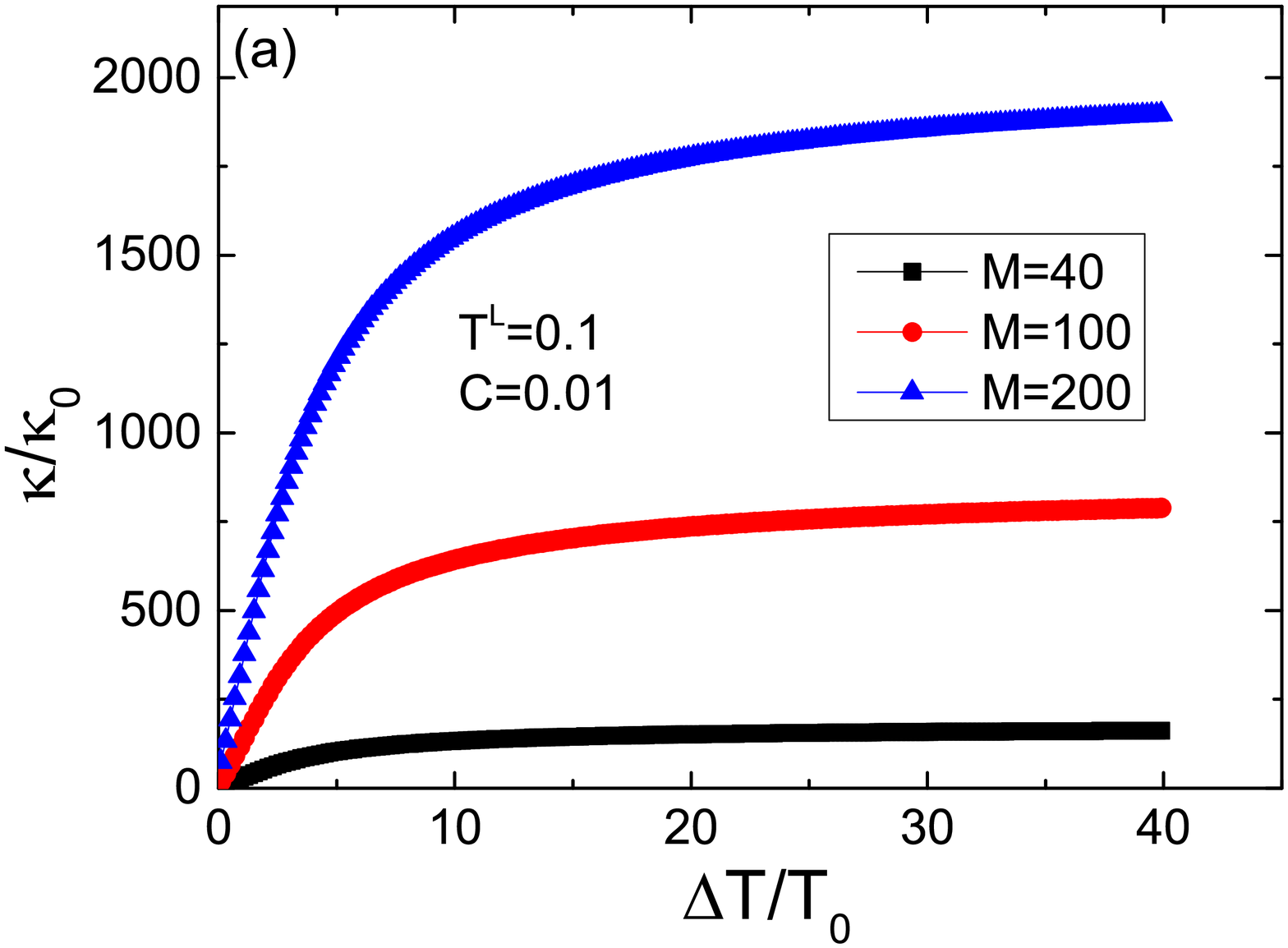}
\includegraphics[width=3in]{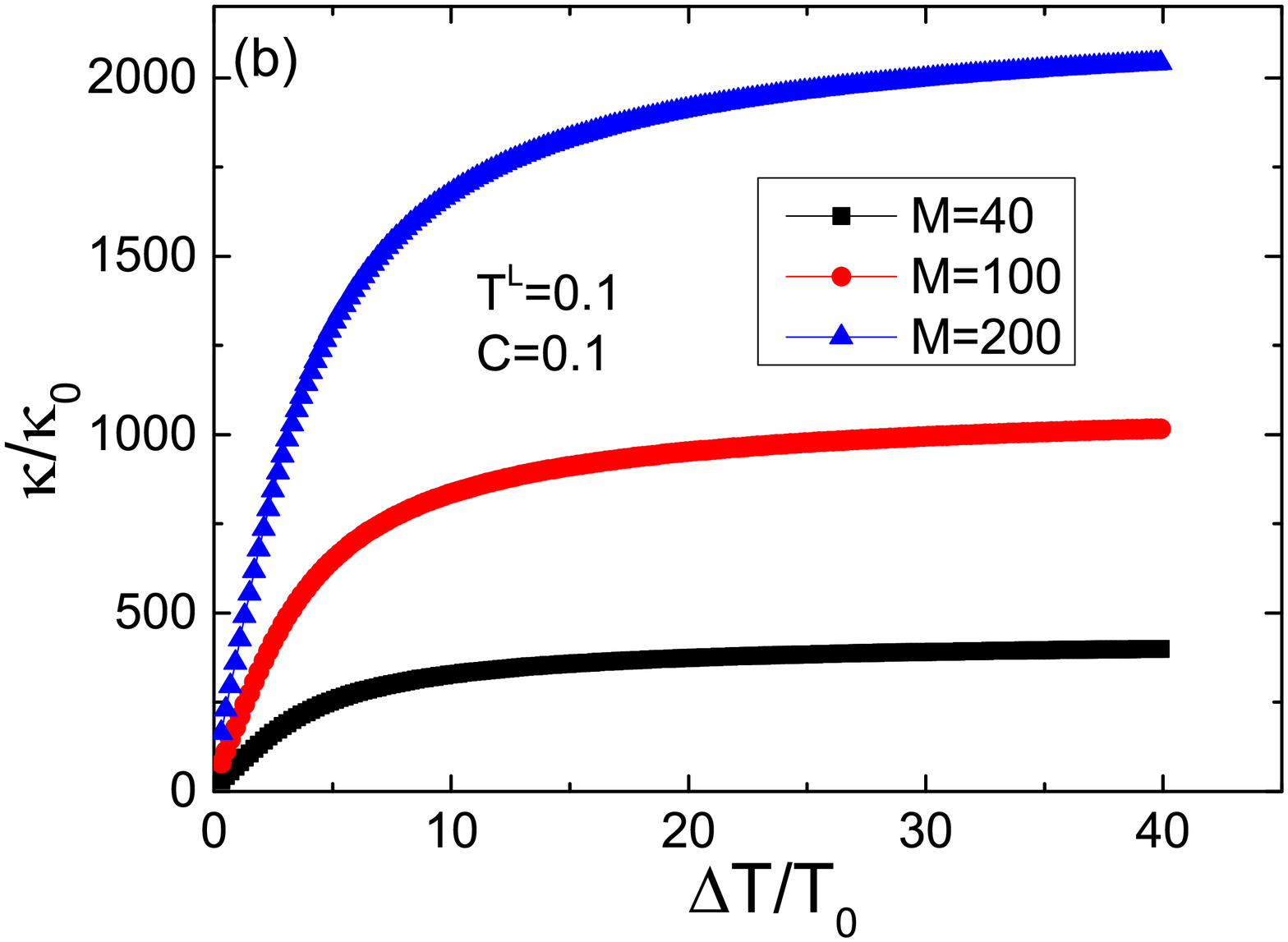}\\
\includegraphics[width=3in]{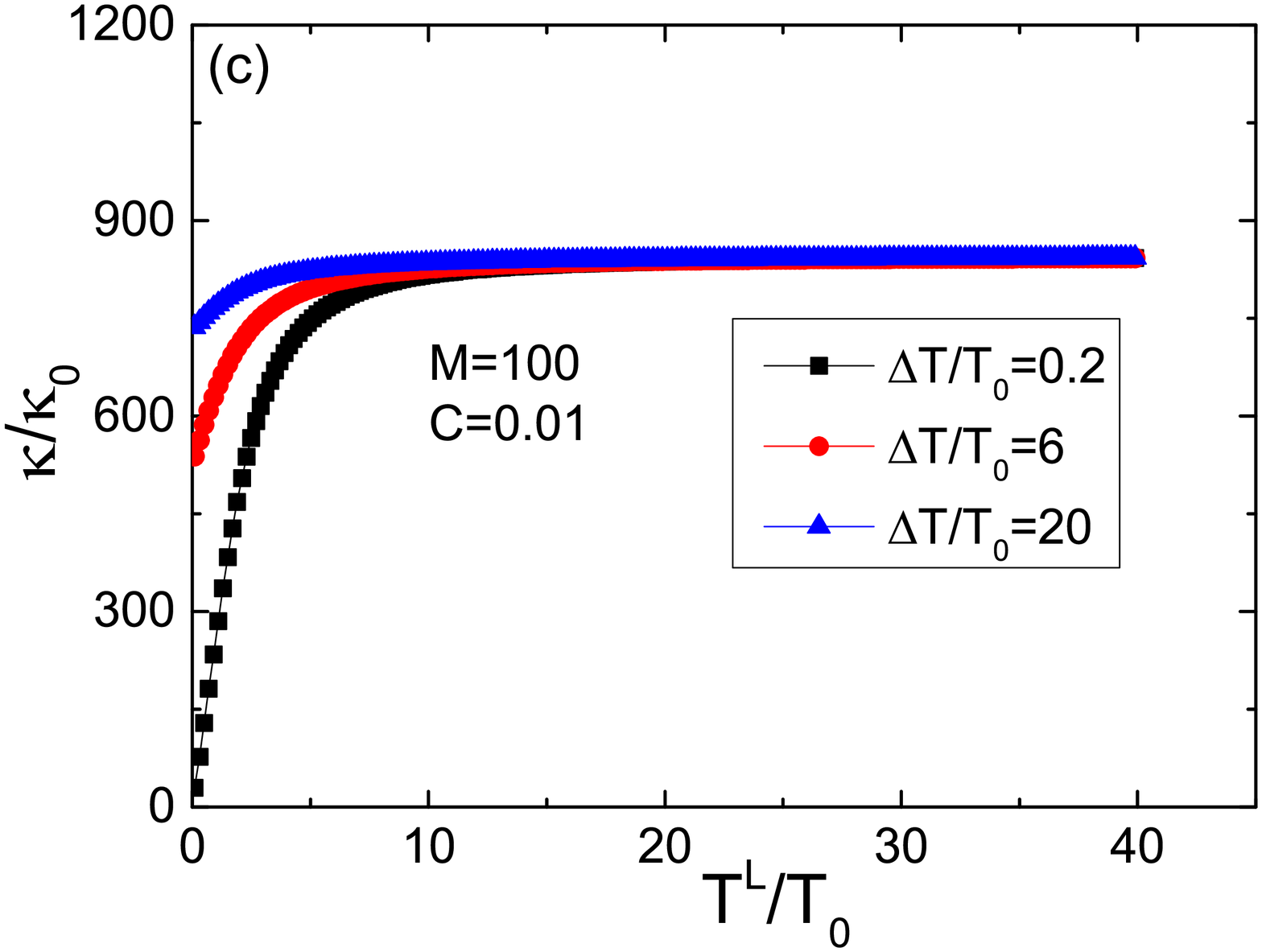}
\includegraphics[width=3in]{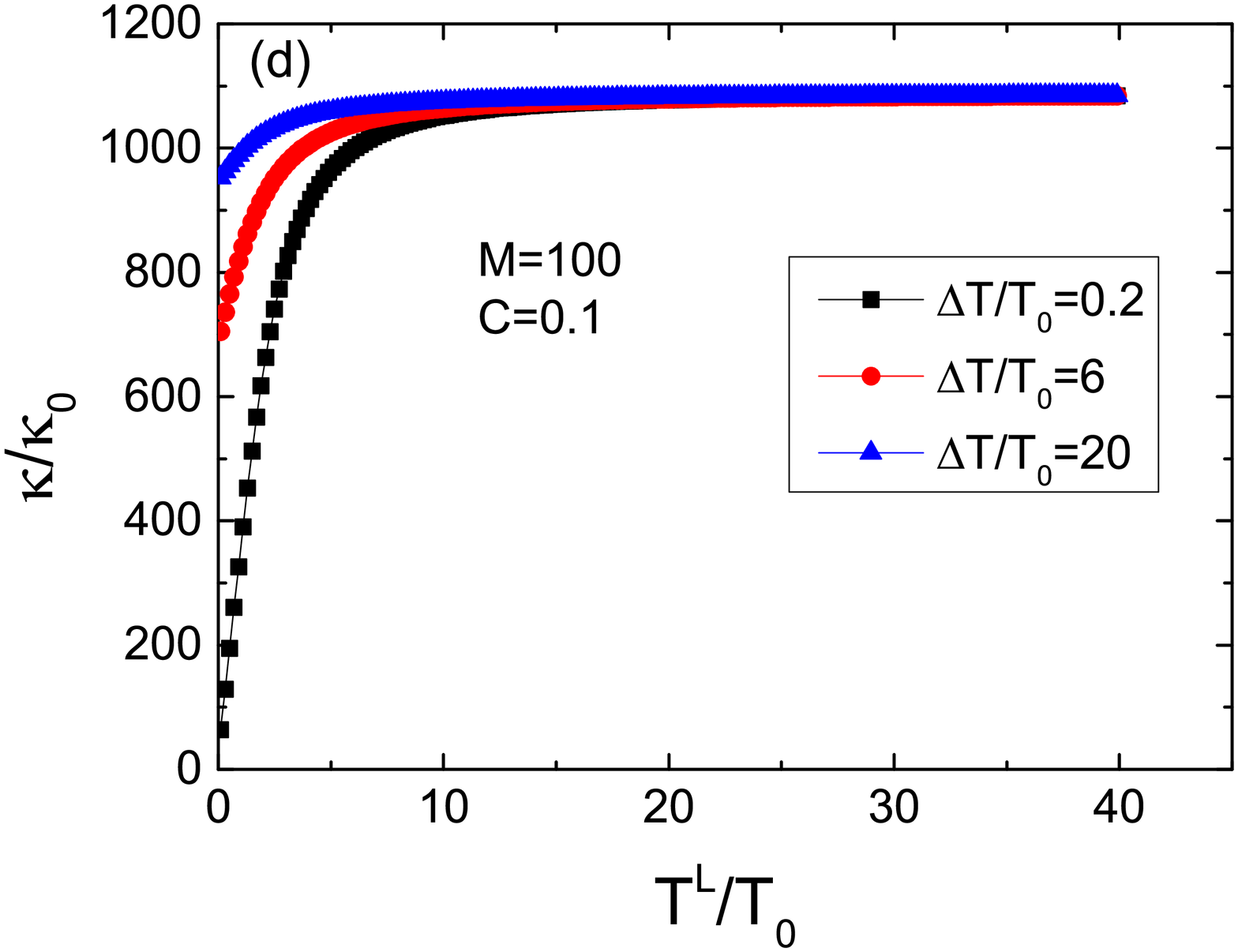}
\caption{nonlinear case for $\kappa$. The $\kappa_0$ and $T_0$ have been defined for dimensionless normalization.(a) the $\kappa$ increases with the increasing of $\Delta T$ with various $M$. (b) the $\kappa$ shows larger values than those in (a) with a larger $C$. (c) the $\kappa$ increases with the increasing of $T^L$ with various $\Delta T$.(d) the $\kappa$ is increased with a large $C$ compared to the results in (c).}
\end{figure}
The dependence of $\kappa$ on $M$ is similar to the results discussed in the linear case. That is, system with a larger $M$ has more phonon channels to get a larger $\kappa$. And strong coupling between the system and the reservoirs can enhance the $\kappa$ by comparing the results in fig.2a and those in fig.2b with two different values of $C$.\\

For a fixed $\Delta T$, the increasing of $T^L$ then enhance the averaged temperature $T$ of the system to get a large $\kappa$, which have been shown in the fig.2c and fig.2d. A larger $\Delta T$ then leads to a larger $\kappa$ by increasing the average temperature $T$ to simulate more phonon channels. But all the $\kappa$ with various $\Delta T$ go to a same constant if $T^L$ is high enough that $\bar{N}(\omega, T)\approx \frac{k_B T}{\hbar \omega}$ is satisfied. Similar to the fig.2a and fig.2b, the comparison between fig.2c and fig.2d shows that a larger $C$ results in a larger $\kappa$ by strengthening the coupling between the system and the reservoirs. 

\section{Conclusions}
We have studied the thermal transport through nanostructures by using the quantum noise theory. The quantum noises are originated from the thermal fluctuation of the reservoirs, with each reservoir in its own thermal equilibrium. A phonon version of quantum Langevin equation has been derived. We apply this QLE to solve the thermal conductivity of the nanostructures. The advantage of this method lays on the fact that we need not to define local temperatures for the system, which actually is in non-equilibrium state.\\

Results show that $\kappa$ of the system is dependent on the phonon channels provided by the system. The oscillator number $N$ and the average temperature $T$ determine the amount of the phonon channels. The phonon transport is also limited by the coupling parameter $C$ between the system and the reservoirs. Larger $C$ makes the phonon transport much easier.      
%\begin{acknowledgements}
%\end{acknowledgements}



\end{document}